\documentclass{emulateapj}

\slugcomment{ver.Jan.21; draft 0pages}
\slugcomment{ver.Feb.12; draft 4pages}
\slugcomment{ver.May 29; draft 11pages}
\slugcomment{ver.Jun 15; draft 14pages}
\slugcomment{ver.Jun 22; 1st circulation}
\slugcomment{ver.Jun 28; revised after Tohru's comments}
\slugcomment{ver.Sep.28; revised after Tetsu's calculation}
\slugcomment{ver.Nov.22; 2nd circulation}
\slugcomment{ver.Dec.14; revised after Kazu/Masami's comments}
\slugcomment{ver.Dec.31; revised after Textcheck}
\slugcomment{ver.Jan.05; 3rd circulation}
\slugcomment{ver.Jan.15; submitted to APJ}
\slugcomment{ver.Mar.20; revised after addressing referee's comments}
\slugcomment{Received 2007 Jan. 15; accepted 2007 Mar. 27}

\shorttitle{LBGs and LAEs around a high-z QSO}
\shortauthors{Kashikawa et al.}

\begin{document}


\title{The Habitat Segregation between Lyman Break Galaxies and Lyman $\alpha$ Emitters around a QSO at $z\sim5$\altaffilmark{1}}

\author{
Nobunari Kashikawa\altaffilmark{2,3},
Tetsu Kitayama\altaffilmark{4},
Mamoru Doi\altaffilmark{5},
Toru Misawa\altaffilmark{6},
Yutaka Komiyama\altaffilmark{2}, 
and
Kazuaki Ota\altaffilmark{7}
}




\altaffiltext{1}{Based on data collected at the Subaru Telescope, which is operated by the National Astronomical Observatory of Japan.}
\altaffiltext{2}{Optical and Infrared Astronomy Division, National Astronomical Observatory, Mitaka, Tokyo 181-8588, Japan; kashik@zone.mtk.nao.ac.jp}
\altaffiltext{3}{Department of Astronomy, School of Science, Graduate University for Advanced Studies, Mitaka, Tokyo 181-8588, Japan.}
\altaffiltext{4}{Department of Physics, Toho University, Funabashi, Chiba 274-8510, Japan.}
\altaffiltext{5}{Institute of Astronomy, University of Tokyo, Mitaka, Tokyo 181-8588, Japan.}
\altaffiltext{6}{Department of Astronomy and Astrophysics, Pennsylvania State University, 525 Davey Laboratory, University Park, PA 16802.}
\altaffiltext{7}{Department of Astronomy, University of Tokyo, Hongo, Tokyo 113-0033, Japan.}

\begin{abstract}
We carried out a target survey for Lyman break galaxies (LBGs) and Lyman $\alpha$ emitters (LAEs) around QSO SDSS J0211-0009 at $z=4.87$.
The deep and wide broadband and narrowband imaging simultaneously revealed the perspective structure of these two high-$z$ populations.
The LBGs without Ly$\alpha$ emission form a filamentary structure including the QSO, while the LAEs are distributed around the QSO but avoid it within a distance of $\sim4.5$Mpc.
On the other hand, we serendipitously discovered a protocluster with a significant concentration of LBGs and LAEs where no strongly UV ionizing source such as a QSO or radio galaxy is known to exist.
In this cluster field, two populations are spatially cross-correlated with each other.
The relative spatial distribution of LAEs to LBGs is in stark contrast between the QSO and the cluster fields.
We also found a weak trend showing that the number counts based on Ly$\alpha$ and UV continuum fluxes of LAEs in the QSO field are slightly lower than in the cluster field, whereas the number counts of LBGs are almost consistent with each other.
The LAEs avoid the nearby region around the QSO where the local UV background radiation could be $\sim100$ times stronger than the average for the epoch.
The clustering segregation between LBGs and LAEs seen in the QSO field could be due to either enhanced early galaxy formation in an overdense environment having caused all the LAEs to evolve into LBGs, or local photoionization due to the strong UV radiation from the QSO effectively causing a deficit in low-mass galaxies like LAEs.
\end{abstract}






\section{Introduction}

%

It is well established that QSOs and/or radio galaxies (RGs) are good markers for galaxy clusters in the local universe \citep{yee87, hal88}. 
Although still very rare, some examples have been definitely established of galaxies strongly clustering around QSOs/RGs even in the high-$z$ universe beyond $z=3$ (e.g., \citealp{zhe06, ove06a, ove06b, ven05, ven04, ven02}), suggesting that such fields are indeed special regions of early galaxy formation.
This trend is naively expected from a generic prediction of essentially every model of structure formation because the luminous objects tend to sit in the local high-density peaks where objects are clustered. 
A search for high-$z$ galaxy clustering around QSOs/RGs thus provides an important test of our basic idea about the biased galaxy formation \citep{kai84}.

In the context of the hierarchical galaxy formation, it is easy to speculate that the merging process triggers the activity of the central massive galaxy/AGN. 
\citet{kau00} asserts that a QSO represents a brief phase in the early life of every galaxy and that the QSO activity is triggered by the major mergers that a galaxy experiences during its assembly.
Such co-evolution of the QSO and galaxy is also suggested by the tight correlation between the mass of the central supermassive black hole and the bulge stellar mass \citep{mag98} or velocity dispersion \citep{geb00}, and that the downsizing trend in the luminosity functions of AGNs and galaxies are similar \citep{ued03}.
The overdense region of galaxies around high-$z$ QSOs, which deserves to be called a protocluster, might be the first site for the co-evolution of QSOs and galaxies.

Beyond $z=3$, the detection of star-forming galaxies basically relies on two major diagnostics in their spectral energy distribution (SED): the redshifted Ly$\alpha$ emission line and the strong Lyman break in the continuum.
Both characteristics have been promising tools for identifying high-$z$ galaxies, called Lyman $\alpha$ emitters (LAEs) and Lyman break galaxies (LBGs), respectively.
LAEs have often been found to be associated with high-$z$ QSOs/RGs or to be tracing large-scale structures, especially at $2<z<3$ (e.g., \citealp{pas96, pen00, kur00, pal04}).
\citet{ven05} found $31$ LAEs around the luminous radio galaxy MRC0316-257 at $z=3.13$.
Their narrowband (NB) imaging and very deep spectroscopy revealed a structure of at least $3.3\times3.3$ Mpc$^2$.
\citet{hu96} discovered two LAEs associated with the QSO BR2237-0607 at $z=4.55$, which was the most distant QSO known at the time.
LBGs have also been found to be clustered around high-$z$ QSOs/RGs.
\citet{zhe06} found an overdensity of galaxies, whose very red color was consistent with that of LBGs, around a radio-loud QSO, SDSS J0836+0054 at $z=5.8$.
The fraction of LAEs among their color-selected red galaxies is expected to be small based on their SED model simulation.
\citet{ove06a} found that LBGs are more strongly clustered than LAEs around the most distant known RG, TN J0924-2201 at $z=5.2$.
\citet{ove06b} confirmed the trend that LBGs are more strongly associated with a RG than LAEs in the case of TN J1338-1942 at $z=4.1$.
On the other hand, some discoveries of high-$z$ protoclusters have occurred in which no QSO/RG activity is observed.
\citet{shi03} discovered a clustering region of LAE candidates at $z=4.86$ in the Subaru Deep Field, which was not a survey targeted for QSOs/RGs.
\citet{ouc05} also discovered two overdensity regions of LAEs at $z=5.7$ in the Subaru/XMM-Newton Deep Field.
Interestingly, no evidence for either QSO/AGN or giant Ly$\alpha$ nebulae is seen in these structures.

By definition, the Lyman break method should detect both LBGs without Ly$\alpha$ emission and LAEs when their continuum fluxes are sufficiently bright.
The physical and/or evolutionary connection between LBGs and LAEs and the cause of the presence or absence of Ly$\alpha$ emission in these high-$z$ star-forming galaxies are still open questions.
\citet{sha01} proposed a plausible scenario of two Ly$\alpha$ bright phases: a galaxy appears as a LAE during its initial starburst epoch when it is still dust free, and then becomes a dusty LBG having Ly$\alpha$ absorption after the ISM metal enrichment.
It becomes less dusty, and hence a LAE, again following the onset of a superwind when it becomes more than a few $10^8$ yr old.
To address the connection between LBGs and LAEs, we must investigate both populations simultaneously at the same place and at the same redshift; a few studies \citep{ste00, ove06a, ove06b} have been conducted with this motivation in mind in high-$z$ protocluster regions.
Galaxies in a high-density environment are likely to start forming earlier than in the general field, and studies of galaxy formation in the field only may have missed possible rare active spots associated with rich protoclusters.


In addition, another unsolved issue about the nature of high-$z$ galaxies is the environmental effects, especially for low-luminosity galaxies.
Low-luminosity galaxies are closely bound up with the surrounding IGM.
Photoionization heating by strong UV background radiation evaporates the collapsed gas in a galactic halo and inhibits gas cooling \citep{bar99}.
This effect is inefficient for bright ($L>L^*$) galaxies residing in deep potential wells, but heavily suppresses star formation in lower mass objects \citep{ben02}.
This could be an important feedback process from AGNs to galaxies.
At $z\sim5$, when the reionization of the universe has completed, the formation of low luminosity galaxies can be suppressed by strong UV background radiation, which will be more intense with the presence of nearby strong UV ionizing sources such as QSOs.
Consequently, the luminosity function of galaxies around a QSO is expected to be flatter than that of the general field.
This problem, which concerns a low-mass population as a ^^ ^^ building block" in the high-$z$ universe, can only be addressed using 8-10m telescopes.

We carried out a targeted search for galaxy clustering in the vicinity of QSO SDSS J0211-0009 at $z\sim5$, using it as a marker for early galaxy formation sites. 
We searched simultaneously for LAE and LBG clusterings around the QSO over a relatively large field of view.
Although the Lyman break method samples a broader range of redshift than NB-imaging, it can be speculated that there is an associated structure with the QSO if any overdensities are found around it (e.g., \citealp{zhe06}).
We found a remarkable contrast between the spatial distribution of LAEs and LBGs in the strong UV background field.

The outline of the paper is as follows.
In \S~2, we describe the observations, data reduction, and LBG/LAE sample selection.
We compare the clustering trends and luminosity functions of the LBG/LAE in fields with and without strong UV background sources in \S~3.
In \S~4, we discuss possible interpretations of the results.

Throughout the paper, we analyze in the flat $\Lambda$CDM model: $\Omega_m=0.3$, $\Omega_\Lambda=0.7$, 
and $H_0=70h_{70}$kms$^{-1}$Mpc$^{-1}$. 
These parameters are consistent with recent CMB constraints \citep{spe06}.
Magnitudes are given in the AB system.

\section{The Data}

\subsection{Observation and Data Reduction}

The target field to search for a protocluster contains QSO SDSS J0211-0009 at $z=4.874$ \citep{fan99}. 
We choose this QSO to coincide its redshifted Ly$\alpha$ emission line with the central wavelength of a NB filter ^^ ^^ $NB711$'' ($\lambda_c=7126$\AA: $\delta\lambda=73$\AA).
The $NB711$ narrowband filter was carefully designed to avoid strong OH night sky emission lines, and was used in the study of field LAEs \citep{shi03}.
The imaging data were taken with Suprime-Cam \citep{miy02} on the Subaru telescope.
Observations were made on the nights of UT $2004$ September and October in Johnson $V$, SDSS $i'$,$z'$, and $NB711$ filters.
Suprime-Cam has ten $2$k $\times 4$k MIT/LL CCDs, and covers a contiguous area of $34'\times27'$ with a pixel scale of $0.''202$ pixel$^{-1}$.
We could sample both LAEs at $z=4.86\pm0.03$ with this NB filter and LBGs at $z\sim5$ with broadband imaging.
The broadband imaging also helped to discriminate LAEs at $z=4.87$ from low-$z$ H$\alpha$, $[$O {\sc iii}$]$ and $[$O {\sc ii}$]$ emitters. 

There is another QSO, SDSS J0210-0018 ($z=4.77$) \citep{fan01} with only a $10.^\prime36$ separation from SDSS J0211-0009.
A single Suprime-Cam's FOV can sample both these QSOs simultaneously, although the redshift of SDSS J0210-0018 is out of the corresponding range of the wavelength coverage of $NB711$.
Thus, we could only sample LBGs in the SDSS J0210-0018 field.
The actual pointing center of the survey field is $02^h 10^m 53.^s0, -00^{\circ} 13^{\prime} 44.^{\prime \prime}4$ (J2000), which is almost the midpoint between the two QSOs.
Our original motivation for targeting the field was to find a coherent large-scale structure of LBGs bridging the two QSOs (e.g., \citealp{fuk04}); however we found no evidence of such a structure.


The integration time was $12000$-$16800$ sec in each band.
For the broadband filters, the typical unit exposure time was $900$ s for $B$ and $240$ s for $z'$; shorter exposures were used for redder filters because of the increase in sky brightness with wavelength.
For the $NB711$ filters, the typical exposure time was $1800$ s.
We adopted a common dithering circle pattern of a full cycle of dithering consisting of $8$ pointings.
The sky condition was fairly good with a seeing size of $0.5$-$0.9$ arcsec.

The data were reduced in almost the same manner as in \citet{kas04},  but we used a semi-automatic reduction system, Distributed Analysis System Hierarchy (DASH) \citep{yag02}.
We performed object detection and photometry by running SExtractor version 2.1.6 \citep{ber96} on the images.
Object detection was made in the $z'$-band and $NB711$-band images for LBG and LAE selections, respectively.
For all objects detected in a given band pass, the magnitudes and several other parameters were measured in the other band passes at exactly the same positions as in the detection-band image, using the {\lq}double image mode{\rq} of SExtractor.
We detected objects that had $6$ connected pixels above $2\sigma$ of the sky background RMS noise and took photometric measurements at the $2\sigma$ level.
The size of the final images was $7330\times 9900$ pixels (corresponding to $24'.7 \times 33'.3$) after removal of very low-$S/N$ regions near the edges of the images.
Object detection and photometry is significantly less efficient and reliable close to very bright stars due to bright halos and saturation trails.
A similar degradation occurs near the edges of the images because of low $S/N$ ratios. 
Unfortunately, the target field contained some bright stars (see Figure~\ref{fig_skyall}).
We carefully defined ^^ ^^ masked regions" corresponding to these low-quality areas, and removed all objects falling within the masked regions.

Aperture photometry was performed with a $2\arcsec$ diameter aperture to derive the colors of the detected objects.
Photometric calibration was made with standard stars in the Landolt field SA 92 \citep{lan92} for the $V$-band magnitudes, spectroscopic standard stars GD50 for the $i'$ and $NB711$-bands, and Feige110 for the $z'$ band.
The target field comprised several stars with SDSS (DR4) identification and photometry, with which we checked the zero-point consistency of the $i'$ and $z'$-band magnitudes.
The $3\sigma$ limiting magnitudes in $2\arcsec$ apertures were $27.84$, $27.18$, $26.13$, and $26.76$ in $V$, $i'$, $z'$, and $NB711$, respectively.
The final co-added image had an effective area of $742$ arcmin$^2$.

\subsection{The LBG sample}

\begin{figure}
\epsscale{1.4}
\plotone{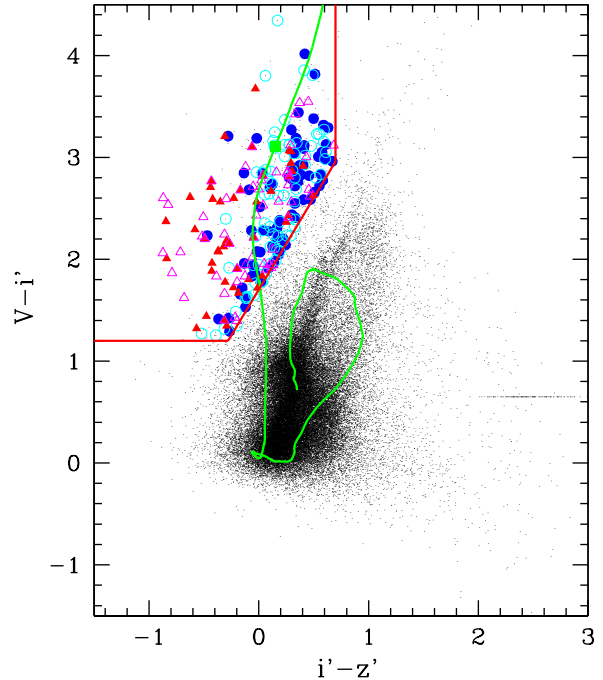}
\caption{$Vi'z'$ color-color diagram used for the LBG selection. 
All the objects detected down to $z_{tot}=26.13$ ($3\sigma$) are plotted.
The objects (above the red lines) in the upper left were defined as LBGs at $z\sim5$.
The (green) solid curve indicates the model SED track (details are described in the text), and the (green) square on the track denotes the position of a LBG at $z=4.87$.
The (blue) solid and (cyan) open circles denote the LAAs in the QSO and cluster fields, and the (red) solid and (magenta) open triangles indicate the LAEs in the QSO and cluster fields, respectively.
\label{fig_lbgsel}}
\epsscale{1.0}
\end{figure}

The LBG sample was constructed using a color selection technique that distinguishes high-$z$ galaxies having a strong Lyman break in their SEDs.
We used the ($V-i'$) versus ($i'-z'$) two-color plane to identify LBGs at $z\simeq5$.
The exact selection criteria of the LBG sample were

\begin{eqnarray}
V-i'  & \geq & 1.2, \nonumber\\
i'-z' & \leq & 0.7, \nonumber\\
V-i'  & \geq & 1.8(i'-z')+1.7.
\label{eq-lbg}
\end{eqnarray}
These criteria were almost the same as \citet{yos06} except that we did not apply the criterion, $B>3\sigma$ (where 3$\sigma$ means a 3$\sigma$ limiting magnitude), which did not greatly affect the sample selection.
Therefore, the completeness and contamination of the LBG sample were expected to be almost the same as evaluated by \citet{yos06}, who estimated the completeness of their LBG sample to be $60\%$---$50\%$ at $z'=24.15$---$26.15$.
According to the estimate by \citet{yos06}, this $Vi'z'$-selected LBG sample had a redshift distribution at $3.7 \lesssim z \lesssim 5.5$ with a peak around $z\sim4.8$ and FWHM $\Delta z=0.8$.
Our LBG criteria were incomplete for red continuum LBGs (i.e., larger $i'-z'$ color) due to dust extinction, while the contamination of galactic stars or low-$z$ galaxies are expected to be negligible ($<20\%$) \citep{yos06}.
Figure~\ref{fig_lbgsel} shows the distribution of objects in the ($i'-z'$) vs ($V-i'$) plane; we also show the evolutionary track of a model galaxy SED.
We used the synthetic galaxy SED templates of Bruzal \& Charlot model \citep{bc03}.
We assumed the Miller \& Scalo IMF from $0.1M_\odot$ to $125M_\odot$, solar metallicity $Z=Z_\odot=0.02$, ages ranging from $10^6$yr to $2\times10^{10}$yr, a constant star formation rate, no dust extinction, and Madau (1995)'s formula for continuum depression due to the intergalactic absorption blueward of Ly$\alpha$ emission.

For cases in which the magnitude of an object was fainter than the $1\sigma$ limiting magnitude, it was replaced with the $1\sigma$ limiting magnitude.
$550$ objects down to $26.13$ mag.($3\sigma$) in $z'_{tot}$, the $z'$-band total magnitude, met our criteria. 
We adopted the {\tt MAG\_AUTO} value derived by SExtractor for the total magnitude.

\subsection{The LAE sample}

\begin{figure}
\epsscale{1.25}
\plotone{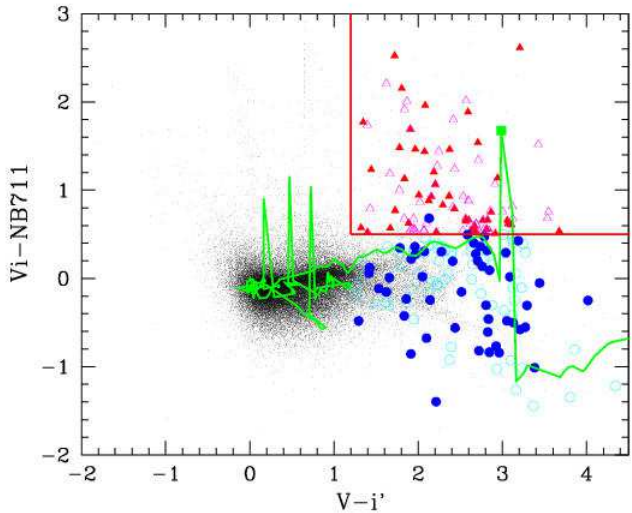}
\caption{$V-i'$ vs $i'-NB711$ diagram used for the LAE selection. 
All the objects detected down to $NB711_{tot}=26.76$ ($3\sigma$) are plotted.
The objects (above the red lines) in the upper right were defined as LAEs at $z=4.87$.
The (green) solid curve indicates the model SED track (details are described in the text) with a rest EW of $100$\AA~as H$\alpha$, $[$O {\sc iii}$]$, $[$O {\sc ii}$]$ emitters, and LAEs, and the (green) square on the track denotes the position of a LAE at $z=4.87$.
The (blue) solid and (cyan) open circles denote the LAAs in the QSO and cluster fields, and the (red) solid and (magenta) open triangles idicate the LAEs in the QSO and cluster fields, respectively.
\label{fig_laesel}}
\epsscale{1.0}
\end{figure}

The LAE sample was constructed using a NB-excess criterion due to Ly$\alpha$ emission and a red color criterion due to the strong Lyman break in their SEDs.
LAEs were selected by the criteria that met both equation (\ref{eq-lbg}) and 

\begin{eqnarray}
Vi - NB711 & > & 0.5,
\end{eqnarray}

where $Vi=V\times0.25+i'\times0.75$ is defined as a rough estimate of the continuum flux at $7126$\AA~wavelength.
The second criterion corresponds to a NB-excess with a rest-frame equivalent width (EW) of $>10$\AA.
Note that the LAE sample contained objects that have a fainter $z'$ magnitude than $26.13$ mag($3\sigma$), although all the LAE objects were detected at above $2\sigma$ in the $i'$-band.
The first criterion as in equation (\ref{eq-lbg}) excludes foreground galaxies with emission lines such as H$\alpha$ at $z=0.09$, $[$O {\sc iii}$]$ at $z=0.42$, $[$O {\sc ii}$]$ at $z=0.91$, and so on.
Figure~\ref{fig_laesel} shows the distribution of objects in the ($V-i'$) vs ($Vi - NB711$) plane as well as the model SED tracks.
We used the same model tracks as in LBG selection, although we added emission line fluxes with a rest equivalent width of $70$\AA~as H$\alpha$, $[$O {\sc iii}$]$, $[$O {\sc ii}$]$ emitters, and LAEs for each corresponding redshift to $NB711$.

As in the case of LBG selection, 
the magnitude was replaced with the $1\sigma$ limiting magnitude when it was fainter than the $1\sigma$ limiting magnitude.
$221$ objects down to $NB711_{tot}=26.76$ ($3\sigma$) met our criteria. 
The detection completeness as a function of $NB711$ magnitude was estimated in the same way as in \citet{kas06b} by counting detected artificial objects distributed on the real $NB711$ image.
The detection completeness was thus found to be $>90\%$ at $NB711<25.5$ and $50\%$ at the limiting magnitude $NB711=26.5$.

\subsection{The LAA and LAE classification}
There is a sample overlap between the LBGs and LAEs that we selected as above.
Some LBGs had Lyman $\alpha$ emission lines that also categorized them as LAEs.
Some LAEs did not meet the criteria of LBGs because their $z'$-band magnitudes were fainter than $3\sigma$ limiting magnitude.

We hereafter distinguish between galaxies with and without Lyman $\alpha$ emission, and we define the ^^ ^^ Lyman $\alpha$ absorbers (LAAs)" as ^^ ^^ Lyman break galaxies at $z\sim5$ with almost no Lyman $\alpha$ emission line (EW$<10$\AA) at $\sim7126$\AA~", and the ^^ ^^ Lyman $\alpha$ emitters (LAEs)" as ^^ ^^ Lyman $\alpha$ emission (EW$\geq 10$\AA) line galaxies at $z=4.86\pm0.03$".
Therefore, the overlapping samples ($89$ objects) satisfying both LBG and LAE criteria were all categorized as LAEs, and these objects were removed from the LAA sample.
Strictly speaking, the definition of a LAA also includes the galaxy population at $z\sim5$ but not at $z=4.86\pm0.03$.
In the following analysis, we cannot distinguish whether we happen to see superposition on the sky, but any overdensities at specific regions, like the QSO environment or the protocluster of LAEs, are speculated to be physically associated with the region.
Future spectroscopy could enable to clearly determine their accurate redshifts; however, most of the LAA/LAE galaxies are too faint to be identified with current spectroscopic capability.

We have $463$ LAA samples down to $z'_{tot}=26.13$ ($3\sigma$), and $221$ LAE samples down to $NB711_{tot}=26.76$ ($3\sigma$).
Based on the redshift distribution of LBGs at $z\sim5$ estimated by \citet{yos06}, the expected probability of falling within the narrow redshift range $z=4.86\pm0.03$ for a LAA sample is $8.8\%$, which gives the expected number of LAAs to have $z=4.86\pm0.03$ to be $\sim63.95$ down to $z'_{tot}=26.0$ after correcting for completeness.
On the other hand, the number of LAEs down to $z'_{tot}=26.0$ is $103.5$ after correcting for detection completeness.
Therefore, the fraction of NB excess objects to all Lyman break selected objects is $61.8\%$, which is larger than the $20-25\%$ evaluated by \citet{ste00} for LBGs at $z=3.1$ down to a similar absolute magnitude $M\simeq-20$.
The LAE fraction of our estimate decreases to $22.6\%$ only when the sample is restricted to a bright LAA/LAE sample with $z'\leq24.5$.
The difference might be caused by either real evolution of the LAE fraction among Lyman break selected galaxies, or by the LAE sample having a fairly high contamination of low-$z$ interlopers.
\citet{shi06} noted from the rest UV LF of the $z=5.7$ LAE sample that $\sim80\%$ or more of the LBG population would have strong Ly$\alpha$ emission at $z\sim6$.


\section{Results}

\subsection{Sky distribution}

\begin{figure}
\includegraphics[scale=0.43,angle=0]{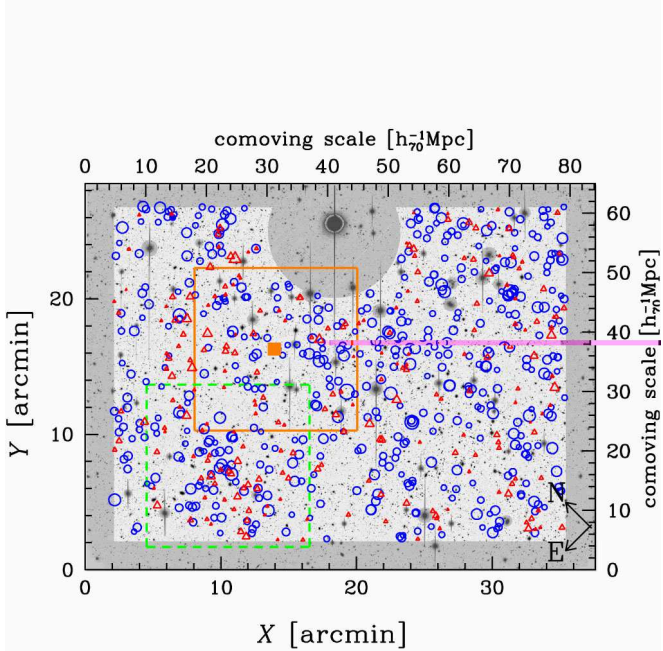}
\caption{The LAA (blue circle) and LAE (red triangles) distributions over the survey field plotted over the $i'$-band image.
The small solid square is the QSO SDSS J0211-0009.
The solid square (orange) is the QSO field, and dashed square (green) is the cluster field.
The shaded regions are the masked regions around bright stars and saturated pixels, in which detected objects were rejected due to low S/N.
There is a very bright star ($mR=9$ mag.) at the top of the figure where we had to mask over a relatively large region because of ghost features, especially in the NB-image.
[{\it See the electronic edition of the Journal for a color version of this figure.}]
\label{fig_skyall}}
\end{figure}

\begin{figure}
\epsscale{1.15}
\plotone{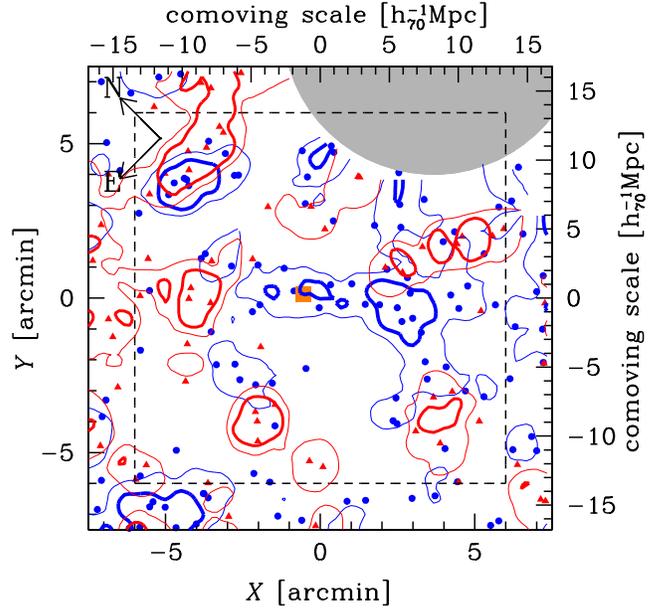}
\caption{The LAA (blue circle)/LAE (red triangle) distribution and its surface number density contours (blue for LAAs and red for LAEs) in the QSO field around SDSS J0211-0009 (central orange square).
The QSO field is defined as a $12\times12$ arcmin$^2$ region, indicated by dashed lines.
The thin and thick contours denote $1\sigma$ and $3\sigma$ excess from the mean local density.
The small mask regions are omitted to avoid complexity, except the largest mask region shown as a shaded area on the upper right corner.
\label{fig_qcont}}
\epsscale{1.0}
\end{figure}

Figure~\ref{fig_skyall} shows the LAA/LAE distributions in the overall survey field.
The distribution of LAEs gives a visual impression to have more inhomogeneous structure than that of LAAs.
We should note that NB searches exploring only a small redshift coverage are sensitive to the large-scale structure, while the LAA distribution is diluted by the sky projection.
We focused on a $12\times12$ arcmin$^2$ ($16\times16 h_{70}^{-1}$Mpc$^2$ in comoving scale) region around SDSS J0211-0009, which we call the ^^ ^^ QSO field".
Figure~\ref{fig_qcont} shows the LAA/LAE distributions in the QSO field.
We also show smoothed surface number density contours for LAAs and LAEs.
The surface number density was measured in a circle of $60\arcsec$ radius with a Gaussian smoothing of $\sigma=10\arcsec$.
The mean value and the dispersion of the surface number density in the circle for each population were derived from the statistics of $10,000$ randomly chosen positions over the entire image.
We confirmed that these mean and dispersion values for LBGs are almost consistent with those measured in a general blank field using the LBG sample at $z\sim5$ \citep{kas06a}, restricting the LBG sample only with $z'<26.31$, the same limiting magnitude as in this study.
The thin and thick contours in Figure~\ref{fig_qcont} denote $1\sigma$ and $3\sigma$ excess from the mean density, respectively.
The distributions of LAAs and LAEs around the QSO were in marked contrast with each other.
The QSO SDSS J0211-0009 lies on a filamentary structure of LAAs extending from the northeast to the southwest, while the LAEs are distributed around the QSO but avoid its vicinity.
Although the Lyman break method samples a broader range of redshifts than NB imaging, the most probable redshift by this $Vi'z'$-color selection is around $z\sim4.8$, almost corresponding to the same redshift of the NB-excess samples.
However, we cannot completely rule out the possibility that the filamentary structure of LAAs around the QSO is attributed to a mere chance of alignment or is incidentally caused by the highly clustered contaminant populations at low-$z$, until spectra are obtained for the LAAs to confirm that they have similar redshifts to the QSO.

\begin{figure}
\epsscale{1.15}
\plotone{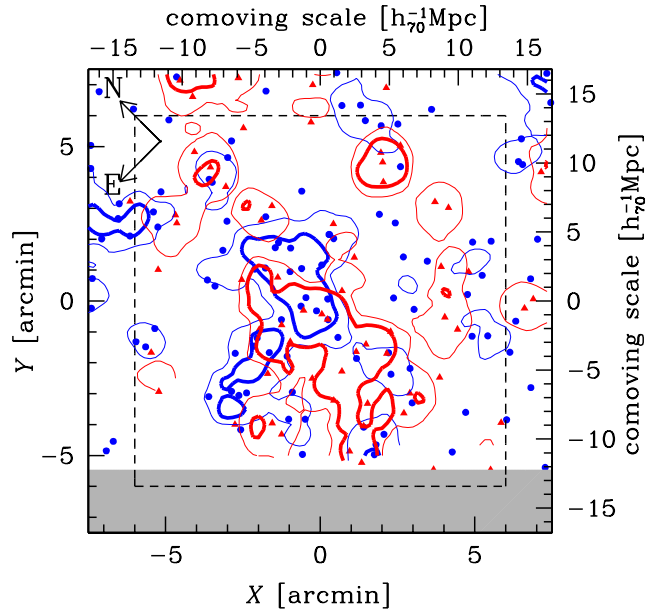}
\caption{Same as Figure~\ref{fig_qcont} but for the cluster field.
The cluster field is within the dashed lines.
The bottom shaded area corresponds to outside the survey field.
\label{fig_ccont}}
\epsscale{1.0}
\end{figure}

We found a galaxy overdensity region $\sim9$ arcmin east of the SDSS J0211-0009 where both LAAs and LAEs apparently have strong clustering.
We hereafter call this region the ^^ ^^ cluster field".
This region has a $6\sigma$ excess of local surface number densities for both LAAs and LAEs.
The projected size of the overdense region is approximately $15h_{70}^{-1}$Mpc$\times20h_{70}^{-1}$Mpc, which is smaller than the $NB$ sampling depth $\sim33h_{70}^{-1}$Mpc.
No QSOs, AGNs, or radio galaxies are known to exist in the cluster field.
Note that the QSO field and the cluster field slightly overlap when an extent of a $12\times12$ arcmin$^2$ is adopted for both fields.
Figure~\ref{fig_ccont} shows the LAA/LAE distributions as well as their smoothed surface number density contours in the cluster field.
In sharp contrast to Figure~\ref{fig_qcont}, the LAAs and LAEs follow a similar distribution in Figure~\ref{fig_ccont}.
They have high number density peaks above $3\sigma$ significance in almost the same position.
The coincidence probability that two independent $\sim2'\times2'$ surface-overdensity regions will overlap within $24'.7 \times 33'.3$ is only about $0.001$.

\begin{figure}
\epsscale{1.35}
\plotone{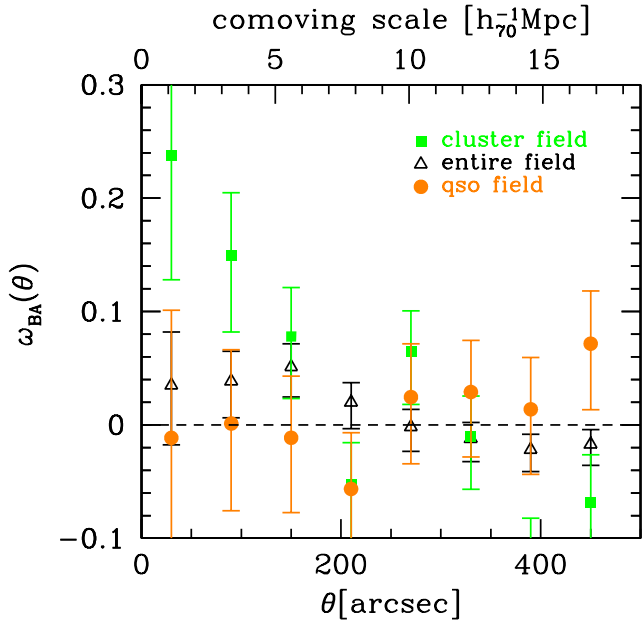}
\caption{The cross-correlation functions between LAAs and LAEs for the entire (black triangle), QSO (orange circle), and cluster (green square) fields.
\label{fig_ccf}}
\epsscale{1.0}
\end{figure}

To observe the trend more clearly, in Figure~\ref{fig_ccf} we show the angular cross-correlation functions (CCFs) $w_{AE}(\theta)$ between LAAs and LAEs for the entire survey field, the QSO field, and the cluster field.
The CCFs were derived using the same estimator by \citet{ade03}
\begin{eqnarray}
w_{AE}(\theta)=\frac{D_AD_E-D_AR-D_ER+RR}{RR},
\end{eqnarray}
where the $D_AD_E$, $D_AR$, $D_ER$, and $RR$ denote the number of LAA-LAE, LAA-random, LAE-random, and random-random pairs having angular separations between $\theta$ and $\theta+\delta\theta$, respectively.
We generated $100,000$ random points to reduce the Poisson noise in random pair counts and normalized $D_AD_E$, $D_AR$, $D_ER$, and $RR$ to the total number of pairs in each pair count.
The random points were created with exactly the same boundary conditions as each field avoiding the mask regions where saturated stars dominate.
The relative amplitude of the CCF among different regions was essential in this study, irrespective of the absolute amplitude of the CCF for each region.
For example in the LAA case, the uncertainty of the CCF due to contamination by foreground galaxies was expected to amount to $20\%$ at most, and we did not correct the incompleteness of the derived CCF in any way.
However, the contamination and completeness should not be much different among the QSO, cluster, and entire fields.
We estimated the Poissonian errors only \citep{ls93} for the CCF as

\begin{eqnarray}
\sigma_w(\theta)=\sqrt\frac{1+w_{AE}(\theta)}{D_AD_E}.
\end{eqnarray}

The CCF in the entire field has a slight amplitude excess at $< 250$arcsec ($\sim9.3 h_{70}^{-1}$Mpc comoving scale), indicating that LAAs and LAEs generally follow nearly identical structures.
The CCF in the cluster field shows a higher amplitude than that of entire field, suggesting that LAAs and LAEs closely coexist in this overdense region.
On the contrary, the CCF in the QSO field shows almost no correlation between LAAs and LAEs on all scales, which matches the visual impression in Figure~\ref{fig_qcont}.

\subsection{Number count}

\begin{figure}
\epsscale{1.35}
\plotone{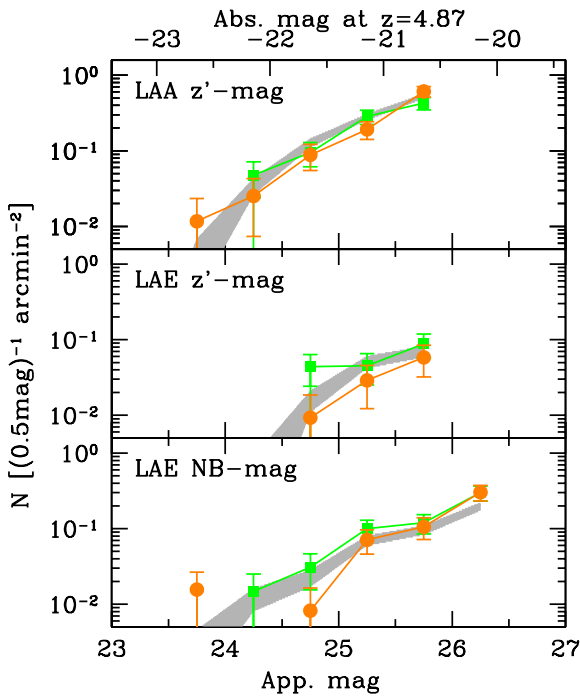}
\caption{The comparisons of number counts between the QSO and cluster fields.
The number counts are investigated with respect to the $z'$-band magnitude for LAAs (upper panel) and LAEs (middle panel), and the $NB711$-band magnitude for LAEs (bottom panel), respectively.
Each number count with $1\sigma$ Poissonian scatter is denoted as a circle (orange) and a square (green) for the QSO and cluster fields, respectively.
The shaded regions show the $\pm1\sigma$ scatter of the number counts for the entire field.
The QSO itself is removed from the number counts.
\label{fig_lf}}
\epsscale{1.0}
\end{figure}

Figure~\ref{fig_lf} shows comparisons of number counts between QSO and cluster fields.
The number counts were investigated with respect to the $z'$-band magnitude corresponding to the rest-frame UV continuum flux ($\sim1540$\AA~) for LAAs and LAEs, and the $NB711$-band magnitude corresponding to the Ly$\alpha$ flux for LAEs, respectively.
The number count for each field is denoted as shaded regions indicating the $\pm1\sigma$ Poissonian scatter of each count.
We corrected the completeness using the estimation by \citet{yos06} for LAAs, and the detection completeness evaluated in \S~2.3 for LAEs, respectively.
We should again note here that no significant difference in the completeness between of the QSO and cluster fields is expected, even if slight uncertainties occur in the estimate procedures.
The black regions show the number counts $\pm1\sigma$ scatter for the entire field, which are regarded as the average number count for each magnitude and each population.
Indeed, the number count on the $z'$-mag for LAAs in the entire field was found to be consistent with \citet{yos06} for a $z'$-band magnitude distribution of field LBGs at $z\sim5$.
The number count on the $NB711$-mag for LAEs in the entire field was also determined to be almost consistent with \citet{ouc03} for a $NB711$-band magnitude distribution of field LAEs at $z=4.86$ and with \citet{shi04} for a $NB704$-band magnitude distribution of field LAEs at $z=4.79$.
Note that the sampling field, $12\times12$ arcmin$^2$, is somewhat larger than the central overdense region in the cluster field, and therefore, the absolute number count in the cluster field might not be significantly higher than the average.

The number counts of the $z'$-mag for the QSO and cluster fields are almost consistent with each other for LAAs, while the $z'$-mag and $NB$-mag counts for LAEs differ slightly between the two fields.
The $z'$-mag number count of LAEs in the cluster field is higher than in the QSO field, although the significance is as small as $1-2\sigma$.
The difference becomes more significant if we take a narrower sampling field for the QSO and cluster fields.
The LAE $NB$-mag number count also shows the same trend, but it is not pronounced.
These differences could be interpreted as that either the number density of LAEs is high in the cluster field and low in the QSO field, or the Ly$\alpha$ (and probably rest-UV continuum) luminosity of LAEs is high in the cluster field and low in the QSO field.
This trend was not seen in the rest-UV magnitudes of the LAA population.
A photometric difference could exist between LAEs in the QSO and cluster fields, although the trend was not sufficiently significant, and the sample in this study was so small that the effect was not firmly established.
We should also note that the discrepancy in the number count of LAEs between the QSO and cluster fields could be caused by the difference of the detection effeciency with $NB$-band for slightly different central redshift in each overdensity.


\section{Discussion}

As seen in \S~3.1, the relative spatial distribution of LAEs and LBGs is in stark contrast between the QSO and the cluster fields.
Why do LAEs avoid the vicinity of the QSO ?
We here consider two possible interpretations of the observational results.
One is that all LAEs have already evolved to LBGs in the overdensity region.
The second is that LAEs cannot form stars due to the photoionization effect of strong radiation from the QSO.

\subsection{Evolutionary sequence from LAEs to LBGs}

\citet{sha01} proposed a plausible scenario of two Ly$\alpha$ bright phases: a galaxy appears as a LAE during its initial starburst epoch when it is still dust free, and then becomes a dusty LBG having Ly$\alpha$ absorption after ISM metal enrichment.
It becomes less dusty, and hence a LAE again, with the onset of a superwind as it gets older than a few $10^8$ yr.
The first phase of this evolutionary sequence is consistent with observational evidence that LAEs are generally younger and lower in stellar mass than LBGs, which have ages of $>10^9$ yr and stellar mass of $>10^{10}M_\odot$ at $z=3-6$ \citep{sha01, sta06, eyl06, yan06}.
\citet{ove06b} found that LAEs around a RG at $z=4.1$ are generally younger and less stellar-massive (a few $\times10^8M_\odot$) than LBGs.
\citet{gaw06} also suggested that field LAEs at $z=3.1$ have a lower stellar mass ($\sim5\times10^8M_\odot$) than LBGs.
The stellar mass of LAEs at a higher redshift, $z\sim5$ \citep{pir06}, was also found to be much smaller ($<10^8M_\odot$) than LBGs.
Recently, \citet{pen07} concluded that LBGs with Ly$\alpha$ emission are much younger and less massive than the LBGs without Ly$\alpha$ emission based on a systematic comparison of spectroscopic sample.
\citet{ven05, dow06} indicate that LAEs have, on average, smaller size than LBGs, on the basis of HST/ACS high-resolution imaging.
The transition from LAEs to LBGs is also suggested in a numerical simulation by \citet{mu06}.
In addition, the tendency for an overdense environment to favor early galaxy formation has been suggested by \citet{gao05}, and thus, an overdense region in a QSO field could contain a more evolved population.
\citet{ste05} showed a direct evidence suggesting that virialization redshift is higher for galaxies in the protocluster region than field galaxies.
Therefore, one interpretation might be that all the galaxies around SDSS J0211-0009 have already passed the first starburst phase as LAEs and are now observed as LBGs forming a filamentary structure around the QSO.

However, some evidence contradicts this interpretation.
First, in the cluster field, which shows a more significant overdensity than the QSO field, LBGs and LAEs coexist.
The interpretation cannot expain the difference of LBG/LAE clustering property in the QSO and the cluster fields.
Second, if LAEs were in previous times clustered around QSOs, some examples might be discovered at $z>4.8$; however, the observational evidence appears to suggest exactly the opposite.
LAEs are usually found clustered around QSOs/RGs at $z<3$ \citep{pas96, pen00, ven05}, while only a few examples \citep{hu96} of them are seen around QSOs/RGs at higher redshifts.
Instead, LBGs are often found around QSOs/RGs at $z>3$.
\citet{ven04} found a LAE overdensity around RG TN J0924-2201 at $z=5.2$, although \citet{ove06a} reported that LBGs are more strongly clustered than LAEs in this region.
\citet{ove06b} confirmed the trend that LBGs are more strongly associated with a RG than LAEs in the case of TN J1338-1942 at $z=4.1$.
This inconsistency between the trend observed and a possible evolutionary sequence from LAEs to LBGs might be caused by observational bias; NB-imaging is difficult for a high-$z$ galaxy survey, which has usually been carried out with HST/ACS.
More studies, especially on estimates of age and stellar mass of LAEs at all epochs and their variation with environment, are required to validate this hypothesis.

\subsection{Photoionization by a QSO}

As discussed above, LAEs are generally revealed to be younger and have less stellar mass than LBGs; however, their dynamical mass has not been clearly determined.  
\citet{ham04} estimated a relatively high halo mass of LAEs at $z=4.86$ of $\sim10^{12} M_\odot$, which was, however, calculated from their clustering strength in a protocluster region.
While \citet{shi06} concluded that LAEs at $z=5.7$ are distributed almost homogeneously in the same survey area.  
Large uncertainties still exist in the estimated dynamical mass of LAEs against the cosmic variance. 
In contrast, the dynamical mass of LBGs at $z\sim5$ has been well determined as $10^{11-12} M_\odot$ based on their clustering strength and the UV luminosity - halo mass relation predicted by semianalytic model combined with high-resolution N-body simulation \citep{kas06a}.

It has been recognized that the formation of low mass galaxies is suppressed in the presence of an external UV field because of photoionization (e.g., \citealp{ui84, rees86, bab92, efs92, tho96}).
If LAEs are generally less massive than LBGs, their formation can be prevented to a greater extent in the vicinity of such radiation sources. 
This effect may explain our finding a different LAE distribution in the QSO field and the cluster field.

We estimated quantitatively the local UV radiation strength around the QSO.  
The local flux density at the Lyman limit, $\nu_L$, at a radius $r$ from the QSO is

\begin{eqnarray}
F^Q_\nu(\nu_L)=\frac{L(\nu_L)}{4\pi r^2},
\end{eqnarray}


where $L(\nu_L)$ is the QSO luminosity at the Lyman limit.
For SDSS J0211-0009, it can be derived by measurements of the continuum AB magnitude in the rest frame $1450$\AA~corrected for Galactic extinction, $AB_{1450}=19.93$, and the continuum slope $\beta=-0.99\pm0.70$ \citep{fan01}, assuming $F^Q_\nu(\nu) \propto \nu^\beta$.
In Figure~\ref{fig_qcont}, LAEs distributed outside $r>770$ physical kpc (corresponds to $>4.5$ comoving Mpc at $z=4.87$).
The local flux density at $\nu_L$ inside a radius of $r=770$ physical kpc was estimated to be

\begin{eqnarray}
F^Q_\nu(\nu_L)=9.0\times10^{-19} {\rm ergs~cm}^{-2} {\rm s}^{-1} {\rm Hz}^{-1}.
\label{eq-fqso}
\end{eqnarray}

This value corresponds to an isotropic UV intensity at the Lyman limit of

\begin{eqnarray}
J_{21}\sim72,
\label{eq-j21}
\end{eqnarray}

when the UV intensity assumes a power-law spectrum of $J(\nu)=J_{21}(\nu/\nu_L)^\alpha\times 10^{-21}$ ergs cm$^{-2}$ s$^{-1}$ Hz$^{-1}$ sr$^{-1}$.
The intensity of the background radiation at $z=4.5$ was evaluated to be log$(J(\nu_L))\sim-21.2$ by the QSO proximity effect measurements of \citet{coo97,wil94}.  
Assuming that the background radiation at $z=4.87$ is almost the same as this value, the relative strength of the local QSO ionizing radiation was estimated to be $\sim100$ times the background radiation at LAEs located $4.5$Mpc distant from the QSO.  
Therefore, the local UV radiation is actually enhanced by the QSO at the position where the deficit of LAEs is observed in the QSO field.

We have investigated how this locally enhanced UV radiation influences the formation of LAEs by means of radiation-hydrodynamic simulations.  
The simulations are an improved version of those described in \citet{kit00, kit01} with an addition of helium atoms in the chemical reaction network.  
Under the assumption of spherical symmetry, they solve self-consistently the dynamics of dark matter and baryons, non-equilibrium chemistry of nine species (e, H, H$^+$, H$^-$, He, He$^+$, He$^{++}$, H$_2$ and H$_2^+$), and the radiative transfer of ionizing photons.  
Self-shielding of H$_2$ against photo-dissociation is also taken into account. 
As an initial density contrast for the dark matter and baryons, we adopt a spherical top-hat perturbation that has just turned around from the Hubble expansion at $z=8.3$ and is destined to collapse (in the absence of thermal pressure) at $z=4.87$. 
We assume for simplicity that the gas is exposed to constant and isotropic external UV radiation with a spectral index $\alpha=-1$, which is close to the observed mean for SDSS J0211-0009. 
In order to cover a wide range of the UV source distance, we vary the UV intensities over the range $J_{21}=0.1 \sim 10^4$.  
Given large uncertainties as to when the radiation source turns on, we first explore the case in which the gas has been exposed to the UV radiation since the turn-around epoch, $z=8.3$; the impact of photoionization is likely to be maximal in this case. 
We will also discuss separately the case in which the radiation source turns on later.

\begin{figure}
\epsscale{1.15}
\plotone{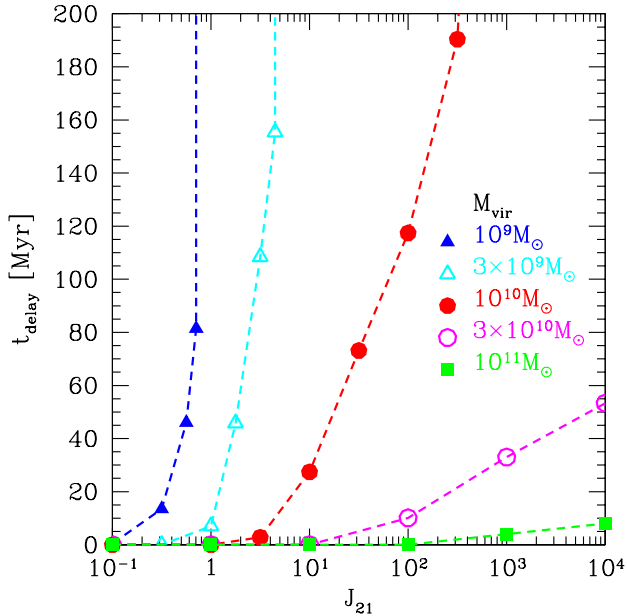}
\caption{
The delay time of star formation, $t_{\rm delay}$, as a function of the radiation intensity, $J_{21}$, for a given virial mass, $M_{\rm vir}$, according to the hydrodynamical model of \citet{kit00, kit01}. 
The delay time is counted from the time when $1/10$ of the baryon mass of each halo was cooled without UV radiation.
The local radiation density in the region where LAEs are deficit around the QSO corresponds to $J_{21} \ga 72$.
\label{fig_model_tj}}
\epsscale{1.0}
\end{figure}

In general, the enhanced UV radiation leads to a considerable delay in star formation in low-mass halos. Figure~\ref{fig_model_tj} illustrates the delay time of star formation as a function of radiation intensity for a given virial mass of a halo, $M_{\rm vir}$. 
The delay time is defined as the time between the epochs at which $1/10$ of the baryon mass in a halo has cooled with and without the external UV radiation. 
The gas is regarded as ``cooled'' when its temperature drops below $1000$K or its density exceeds $10^6$cm$^{-3}$.
Figure~\ref{fig_model_tj} shows that star formation is completely suppressed in a halo with $M_{\rm vir}<10^{9} M_\odot$ under the background UV field with $J_{21} \sim 1$. 
In contrast, it remains almost unaffected in a massive halo with $M_{\rm vir}>10^{11} M_\odot$ even in the vicinity of radiation sources. 
This is mainly because photoionization can raise the gas temperature only to several times $10^4$ K. 
In an intermediate mass halo, the star formation is suppressed during the initial stage of the collapse and the delay time depends sensitively on the radiation strength. 
For example, the UV intensity corresponding to equation (\ref{eq-j21}) will suppress the star formation for $\sim 10^8$yr in a halo with $M_{\rm vir}=10^{10} M_\odot$. 
Such a halo will be able to host bright galaxies $\sim 10^8$yr later than those placed in the weaker UV field.

The above results imply that if the observed LAE deficiency around the QSO is due to photoionization and the age of the LAE is $10^6 - 10^8$ yr \citep{gaw06, ove06b, pir06}, the LAE mass is $M_{\rm vir} \sim 10^{10}M_\odot$.
This is somewhat smaller than previous observational estimates \citep{ham04}. 
As mentioned already, the dynamical masses of the LAEs are still poorly determined from observations and we do not consider this difference to be critical.
Our results further suggest that photoionization cannot delay the star formation in massive LBGs with $M_{\rm vir}\sim10^{11-12} M_\odot$.

Note that our model is restricted to the case in which stars form at the center of a nearly spherical halo. If star formation takes place after a disk-like collapse or in substructures (e.g., \citealp{su00, mu06}), the impact of photoionization will be greater and the inferred mass of the host halo can be larger.




Another important aspect of the photoionization efficiency by a local ionizing source is the time difference between the QSO activity phase and collapse time of galaxies.  
Even low-mass galaxies could collapse if they form before the QSO activity has grown.  
To examine this possibility, we also performed runs in which halos that have already collapsed at $z=4.87$ with a gas density profile of $\propto r^{-2}$ are exposed to an external UV field.  
In this case, star formation cannot be delayed even in a halo as small as $M_{\rm vir}=10^{9} M_\odot$.
Hence, another possible interpretation of the observational result is that LAAs have already formed, or are in the process of formation, at the time of the active phase of the QSO host galaxy, whereas LAEs are
young, having formed much later than the QSO/LAA formation.  
This scenario is supported by previous results that the stellar mass of LAEs is smaller than LBGs and is almost unchanged at all the epochs \citep{ove06b, gaw06, pir06}, although more accurate estimates of the stellar mass of high-$z$ LAEs should be accumulated in the future.  
This picture is also consistent with the idea that a galaxy first appears as a LAE during its initial dust-free starburst phase \citep{sha01}.  
This interpretation also requires accurate estimates of the formation epochs of LAAs/LAEs relative to the evolutionary timescale of QSO activity, which have not yet been determined.  
\citet{kaw03} proposed a QSO evolution model with a prediction of a timescale for a QSO active phase to be $\sim$ a few $10^8$ yr.
\citet{she07} estimated the timescale of the luminous accreation phase to be $\sim$ a few $10^7 - 10^8$yr for QSOs at $z>3.5$.
These timescales are comparable to the LAE age $\sim10^8$ yr \citep{gaw06} and is shorter than the LBG age $\gtrsim10^9$ yr \citep{sha01} estimated by multicolor SED fitting.  

\citet{ove06b} concluded that LAEs prefer regions that are devoid of UV-bright LBGs in their protocluster region at $z=4.1$.
Their result is inconsistent with ours in the cluster field, where we found that LBGs and LAEs coexist in an overdense region.
The LAEs in their study might not have avoided LBGs but rather the much brighter central radio galaxy, which has a UV luminosity of $6L^*_{z=4}$, although the photoionization effect was less significant than for QSOs.

The importance of radiative feedback from a local strong UV ionizing source is suggested by several lines of observational evidence.
The QSO proximity effect emerged in nearby QSOs, where the relative strength of the local QSO ionizing radiation is only $>0.1$ times the background (e.g., \citealp{baj88, lu91}), which is much more sensitive to the QSO intensity than we found in this study.
\citet{ade03} found that LBGs are associated with H {\sc i} underdensities at a separation of $<0.5h^{-1}$Mpc, based on the statistics of several LBG samples and IGM absorption lines in the spectra of background QSOs.
This result was interpreted as being caused by either photoionization from LBGs or by the LBG galactic wind.
\citet{fra04} detected no Ly$\alpha$ emission within $1$Mpc of a luminous QSO at $z=2.2$, and the neutral hydrogen clouds near the QSO may have been photoevaporated.
Along with these examples, the observational result suggests that the photoionization effect from a local strong UV ionizing source could play an important role in making a possible ^^ ^^ habitat segregation" between LAAs and LAEs in the high-$z$ universe.

\section{Conclusions}

In this paper, we have presented a possible clustering segregation between LAAs and LAEs around a high-$z$ QSO.
We carried out a wide-field target survey for LAAs and LAEs simultaneously around a QSO SDSS J0211-0009 at $z=4.87$.
Our main conclusions can be summarized as follows:

1. LAAs form a filamentary structure including the QSO, while LAEs distribute around the QSO but avoid its vicinity.

2. LAAs and LAEs are spatially cross-correlated in a protocluster field where no strongly UV ionizing source such as a QSO or radio galaxy is known to exist.

3. We found a weak trend that number counts based on Ly$\alpha$ and UV continuum fluxes of LAEs in the QSO field are slightly lower than in the cluster field, whereas number counts of LAAs are almost consistent with each other.

4. The LAEs avoid the region around the QSO, where the UV background radiation could be $100$ times greater than the average intensity at the epoch.
The clustering segregation between LAAs and LAEs seen in the QSO field might be caused by either all the LAEs having already evolved into LAAs in the QSO field, or photoionization around the strong UV source effectively causing a dearth of low-mass galaxies such as LAEs.

These results can only be achieved by wide-field imaging targeting QSO/RG fields.
For example, in the J0211-0009 field, small FOV imaging revealed only the filamentary structure of LAAs around the QSO, without showing the larger surrounding structure formed by LAEs (it ^^ ^^ fails to see the forest for the trees").
Recently, similar large FOV imaging revealed a cosmic web of LBGs around the RG TN J1338-1942 at $z=4.1$ \citep{int06}.
Although our sample cannot, in isolation, allow the derivation of a general picture of the properties of high-$z$ galaxies in an overdense field around known high-$z$ objects, it provides a potentially important perspective for understanding the nature of high-$z$ galaxies in different environments.
Future comparative studies of different fields and different UV background fluxes (different QSO/RG luminosity) would permit more quantitative discussions.

\acknowledgments

We acknowledge Hisanori Furusawa for his helpful support of the observations.
We are very grateful to Kazuhiro Shimasaku and Masami Ouchi for kindly providing the data of their LAE sample and for their valuable comments on the manuscript.
We thank the referee for helpful comments that improved the manuscript.
The research was supported by the Japan Society for the Promotion of Science through Grant-in-Aid for Scientific Research 16740118.

\end{document}